# Expecting the unexpected in the search for extraterrestrial life

Peter Vickers, University of Durham, UK

---

## Abstract

On p. 10 of the 2018 National Academies Exoplanet Science Strategy document (NASEM 2018), 'Expect the unexpected' is described as a general principle of the exoplanet field. But for the next 150 pages this principle is apparently forgotten, as strategy decisions are repeatedly put forward based on our expectations. This paper explores what exactly it might mean to 'expect the unexpected', and how this could possibly be achieved by the space science community. An analogy with financial investment strategies is considered, where a balanced portfolio of low/medium/high-risk investments is recommended. Whilst this kind of strategy would certainly be advisable in many scientific contexts (past and present), in certain contexts – especially exploratory science – a significant disanalogy needs to be factored in: financial investors cannot choose low risk high reward investments, but sometimes scientists can. The existence of low risk high impact projects in cutting-edge space science significantly reduces the warrant for investing in high risk projects, at least in the short term. However, high risk proposals need to be fairly judged alongside medium and low risk proposals, factoring in both the degree of possible reward and the expected cost of the project. Attitudes towards high risk high impact projects within NASA since 2009 are critically analyzed.

## 1. "There's a tension in the 2018 Exoplanet Science Strategy document"

On p. 10 of the 2018 National Academies Exoplanet Science Strategy document (NASEM 2018), 'Expect the unexpected' is described as a general principle of the exoplanet field. In the Exoplanet Science Strategy public briefing which accompanied the publication of this document, Scott Gaudi – co-chair of the strategy committee – reiterates: "One thing I am quite sure of, now having spent more than twenty years in this field of exoplanets … expect the unexpected." This is motivated by the fact that there has been an "avalanche of unexpected discoveries" (*op. cit.*, p. 1) in the field in the past twenty years, and this trend is expected to continue. Gaudi also likes to say, 'Mother Nature is more imaginative than we are' (Goldsmith 2018, Prologue). The implication is that we must remain open-minded about what we will find out there in the universe.

It is easy to say these things, but not so easy to adopt scientific strategies that do justice to them. Reading through the Exoplanet Science Strategy document, it is hard to see how the statement on p. 10 – 'expect the unexpected' – influences anything that comes thereafter. The document is (as one might well expect) saturated with strategy decisions based very firmly on our theoretical expectations. Questions are repeatedly raised concerning such issues as the types of star most likely to host planets with atmospheres, the conditions determining whether planets will be tidally locked to their host star, the spectral signatures most likely to be indicative of life, and so on. Our answers to questions such as these directly influence high-stakes strategy decisions concerning which (types of) telescope to build, which instruments to put on our satellites, which stars to focus our attention on, which planets to rank highest in our priorities for extended study, and so on. And in each and every case our theoretical expectations provide the driving force. When we later encounter paragraphs explicitly putting weight on our 'expectations' (e.g. p. 50), the phrase 'expect the unexpected' seems long forgotten.

One obvious response to this is that we should be concerned: we are not 'expecting the unexpected' when our exoplanet science strategies are steeped in our expectations, and this might have serious consequences for progress: we are liable to miss things precisely because they fall outside our expectations. Well-known psychology experiments demonstrate the difficulty we sometimes have when it comes to 'the unexpected'. In the famous 'gorilla in the room' experiment (Simons and Chabris 1999), roughly 50% of people miss the person dressed in a gorilla costume when their attention is taken up by something else (counting bounces of a basketball). Other experiments in psychology lead to similar findings. Subjects faced with anomalous playing cards, such as a red four of spades, fail to recognise that they are anomalous, putting them into pre-existing conceptual boxes (Bruner and Postman 1949). In other words, they see what they expect instead of what is actually presented to them.

If these examples seem too far removed from scientific practice, there are similar lessons throughout the history of science. The ozone hole was originally missed because ozone readings below a certain level were not taken seriously (Oreskes and Conway 2010, p. 119ff.). In work closely related to this paper, Cleland (2019a; 2019b, Ch. 8) provides several examples from the biological sciences illustrating this general phenomenon, where "assumptions can blind biologists to the significance of puzzling empirical findings" (Cleland 2019b, p. 182). Her first example (2019a, p. 723f) concerns how the discovery of ribozymes was delayed by almost a year and a half, owing to the fact that their possibility departed significantly from expectations. Cleland (2019b, p. 178) usefully clarifies: "It is not so much that such beliefs alter our actual perceptions as that our expectations about a phenomenon, coupled with limitations of our instruments, influence how closely we attend to it."[1]

Exoplanet science is hardly immune to this phenomenon: Mayor and Queloz (1995) discovered the first exoplanet orbiting a sun-like star (51 Peg b) in 1995, but others could have spotted it (or a similar planet) sooner if only they had been inclined to check their data for a large planet with an unthinkably short orbital period – 51 Peg b orbits its star in four days. Gordon Walker's group had used the same method of detection (the radial velocity/wobble method) as Mayor and Queloz for twelve years, between 1980 and 1992, but as Boss (2010, p. 21) notes, "Walker did not search his data for such a short-period orbit. The shortest orbital period he considered was 40 days, which must have seemed ridiculously short at the time." Walker (2003, p. 318) himself distinguishes two separate problems: "we had all been looking in the wrong place," and "our programs had been ill designed to sample such periods." At the same time (late 80s/early 90s) Marcy and Butler were also using this method to search for planets, and fell into the same trap as Walker. A Jupiter-mass planet in a four-day orbit was "absurd" (Marcy and Butler 1998, p. 67). They admit, in hindsight, that they were misled by their expectations, subject to "detection biases" (p. 60). Indeed, when Mayor and Queloz announced the discovery of 51 Peg b in 1995, Marcy and Butler quickly realised that they already had in hand data corresponding to two further planets (70 Vir b, and 47 Uma b). They could find the signatures of these planets in their data only when they knew what they were looking for.

---

[1] Like me, Cleland (2019a; 2019b, Ch. 8) worries about situations where powerful theoretical biases mean that scientists miss something significant; see Section 7, below, for discussion.

Such examples support Hacking's (1983, p. 179) claim that "noticing is theory-loaded,", and perhaps even Kuhn's (1970, p. 24) stronger claim that "those [phenomena] that will not fit the box are often not seen at all." If this is right, the status quo is problematic, because we have good grounds for thinking we are missing things and will continue to miss things: there could be significant phenomena we will miss for decades, so long as our theories provide us with expectations (biases) which serve to block them from our view, or which lead us to dismiss them too quickly as 'uninteresting'.

## 2. "No there isn't – nothing to worry about"

But perhaps these concerns are not really warranted. Scientific communities are diverse, with different types of individual, different teams, different values, different talents, and so on. The 'gorilla in the room' is not missed by *everybody*, because some individuals suffer from so-called 'inattentional blindness' less than others. In science, we actually only need *one* scientist to see something unexpected, and, if it really is significant, very soon the whole community will know about it. In other words, even if 99.99% of the scientific community are somehow 'blind' to something, it still doesn't get missed by that community. In addition, individual scientific teams that originally miss something often correct their mistake after a relatively short delay (in the grand scheme of things). Cleland's own example of the discovery of ribozymes involved a 'delay' of less than 18 months, for example – and it is no delay at all if one thinks that scientific teams *should* explore options in line with expectations before they explore unorthodox options.[2]

Consider also the specific examples that motivated Gaudi to say 'expect the unexpected'. Arguably, these unexpected phenomena were never going to be missed, at least not for long. Drawing on NASEM (2018) and the corresponding public briefing, three examples motivating Gaudi's remark are:

(i)     the abundance of a type of planet (super-Earths/mini-Neptunes) in our galaxy that doesn't feature in our own Solar System (e.g. COROT-7b);

(ii)    planets that are hotter than some stars (e.g. KELT-9b);

(iii)   planets found in the 'wrong' place, given our theoretical expectations of where they would have originally formed (e.g. 51 Peg b).

These were all highly unexpected, and yet as soon as we had the technology such phenomena were (quite obviously) always going to reveal themselves to us. Thus the mantra of the inattentional blindness literature, "we often miss what we don't expect to see", just doesn't apply.

One reason for this is that we design our missions and instruments in a way that allows for 'the unexpected'. The Hubble space telescope, for example, has observed many thousands of unexpected phenomena, including phenomena (including exoplanet phenomena) that formed absolutely no part of the theoretical landscape when the telescope was designed. The Kepler mission was similarly open-minded to what it might find; it focused indiscriminately on a rather random portion of the night sky, continuously observing 150,000 stars, looking for exoplanet transits. It wasn't somehow restricted to only notice phenomena that are expected. And other missions (e.g. WFIRST) will similarly be 'open' to what they will find. Indeed, one might argue that the NASEM (2018) strategy document *does* accommodate 'expecting the unexpected', for example when it stresses the importance of mission flexibility in both design and operation (p. 83).

---

[2] Not that Cleland wants scientists to quickly explore unorthodox options, of course. She just wants the (astrobiology) community to be more *open* to theoretical change; she wants to reduce the "grip of theoretical commitments on the scientific mind" so as to "expedite the recognition of anomalies" (2019b, p. 176). See Section 7, below.

What of the 'theory ladenness of observation' familiar to philosophers of science (e.g. Psillos 1999, p. 78)? A case can be made that it has often been exaggerated. It cannot be said that the theories we have affect everything we see; if that were true, it would be hard to make sense of the fact that scientists just do see phenomena that seriously challenge their preferred theories (cf. Cleland 2019b, p. 177). Indeed, many scientists of the past have been primarily talented as *investigators of phenomena*, adopting a method of investigation that is theory-neutral to a very significant degree. Hacking (1983, p. 157ff.) makes this point using David Brewster as an example. He writes, "Brewster firmly held the 'wrong' theory while creating the experimental phenomena that we can understand only with the 'right' theory, the very theory that he vociferously rejected," and emphasises that Brewster was merely "trying to find out how light behaves." (*ibid*.). Brewster was very successful, discovering new facts about light including birefringence in bodies under stress. His observations of light behaviour were *not* affected by the theory of light that he explicitly preferred (the Newtonian, corpuscular theory).

This all suggests that we will continue to uncover surprises in the future, just as we have in the past, without needing to change our scientific methods in any way.

## 3. Two ways 'the unexpected' might get missed

The fact that we've uncovered unexpected phenomena in the past (the ozone hole; ribozymes; Gaudi's three examples) doesn't go far to supporting the conclusion that we will *always* uncover unexpected phenomena. And the fact that telescope concepts (e.g. Hubble and Kepler) are somewhat 'open-minded' vis-à-vis what they will find, and have uncovered some unexpected phenomena, doesn't go far to showing that we are not missing some/many other important phenomena. Two ways in which the space science community might miss something stand out.

First, consider the competition for time when it comes to our best satellites. The NASEM strategy document states the issue clearly in a discussion of JWST, the James Webb Space Telescope:

> "[W]ith a limited lifetime shared-resource facility there is a trade-off between characterizing many easy-to-observe hot giant planets versus using substantial observing time to perform detailed characterization of a much smaller number of terrestrial exoplanets […]. With hundreds of high-quality atmospheric characterization targets to choose from, multiple choices of observing modes and wavelength coverage, and many competing research groups that have spent years eagerly awaiting the launch of JWST, one might expect an onslaught of observing proposals in the early cycles of the JWST mission. (NASEM 2018, p. 86)

In addition, only ~20% of JWST time will be devoted to exoplanet science (*ibid.*), and the teams that built JWST's instruments as well as select members of the Science Working Group get a certain amount of guaranteed time. The competition for the time left over is incredibly intense.

One thing is trivial: nobody doubts that many things will get missed for a long time, simply because there are too many things to look at. In the past ten years more than 3650 exoplanets have been confirmed – more than one a day. With missions such as TESS we expect this trend to continue (Huang *et al.*, 2018). Perhaps the best we can do is select the most exciting targets (e.g. planets in the habitable zone of the TRAPPIST-1 system; Gillon *et al.*, 2017), and move on to other interesting targets as and when time allows, always keeping an eye on the newly confirmed planets coming out of the raw data.

But how do we decide which targets are the highest priorities? The NASEM (2018) strategy document notes that we must try to "rank exoplanet targets" (p. 51). How will such a ranking be achieved? Quite clearly, our theoretical expectations will play a central role in determining any ranking. For example, if our highest priority is to find extraterrestrial life, and we characterise the habitable zone in a *generous* way, certain planets will be ranked higher than they would with a more conservative habitable zone (cf. Joshi and Haberle 2012). And we should absolutely expect our theoretical expectations to change in all sorts of ways over the next twenty years (say). When it comes to the "avalanche of unexpected discoveries" in the past twenty years, we shouldn't be thinking only about concrete empirical surprises (e.g. the abundance of super-Earths/mini-Neptunes in the galaxy). We should also be thinking about very significant *theoretical* surprises.

Theoretical surprises in the field of exoplanet science have been numerous in the past twenty years, some spurred by empirical discoveries and others a result of relatively 'pure' theoretical research. An obvious theoretical reassessment was demanded by the fact that planets have been found in the 'wrong' place; we now think that planet migration is common (not rare), and models of such migration are a major part of theory. Other significant theoretical turnarounds include: (i) a reassessment of the habitability of terrestrial planets orbiting in the habitable zone of dwarf stars (e.g. Luger and Barnes 2015), and (ii) a reassessment of various suggested 'biosignatures' (e.g. Schwieterman *et al*. 2016).[3] Many other examples could be mentioned; e.g. NASEM (2018, p. 48f.): "It was previously thought that cores of 10 $M_E$ were necessary to initiate runaway gas accretion...". Nobody doubts that such theoretical turnarounds will continue, and this in turn will affect any priority ranking of exoplanet targets. Thus we expect that at least *some* planets ranked low down any proposed list right now will turn out to be highly significant. We could see this significance quite soon, were we to point our best telescopes at them. But as things are, we won't point our best telescopes at them any time soon. And in fact, some such planets will probably *never* get looked at: newly discovered planets will keep being slotted into the top 20% of our ranking, meaning that we will never get around to analysing planets in the bottom 80% of our ranking with our best instruments. Not unless theory changes, and tells us to bump some planets up the list.

---

[3] The latter biosignature case is an example driven by relatively 'pure' theoretical research. New models of exoplanet atmospheres occasionally show that what everyone *thought* would count as a solid 'biosignature' can actually emerge abiotically. Cleland (2019b, p. 186) discusses such a case: "Until fairly recently it was widely believed that the only material systems that can be far out of thermochemical redox equilibrium were biological...". See also Cleland (2019a, p. 725).

Can we shrug our shoulders and say that, if exciting planets are far down the list such that they'll never get looked at, we just need to trust in theory, and wait for the theoreticians to tell us why those planets really should be high-priority targets? But consider, for example, the seasonal variation of methane and oxygen levels on Mars (Moores *et al*. 2019): theoreticians were (very probably) never going to predict this phenomenon from pure theory, leading us to look for it. We had to discover it empirically, and only then did theoreticians get busy developing relevant theory.[4] Similarly, we should expect that at least some planets down the list will be highly significant for theoretical reasons not yet in our possession, and which won't *be* in our possession unless we get empirical data to spur on the relevant theory.

This is *one* way in which we should expect to miss the unexpected: planets currently not expected to be especially interesting targets, given current theory, will get missed because we'll never look at them. We'd look at them if we had different theoretical expectations, but we may never develop the relevant theory *unless we look at them* (a chicken-and-egg situation). The relationship between the discovery of 51 Peg b and the development of planet migration theory is just such a case: the theoretical development is a consequence of the discovery, and the discovery would have come much sooner had we already had the theory. In another such case the discovery might never come without the relevant theory in hand.

There is also a second way in which we might miss the unexpected, having to do with which projects we choose to fund in the first place. What type of project is likely to get funded? An important consideration has to do with the expected payoff for the investment of resources, especially in a worst-case-scenario. For example, missions such as TESS get funded because, even if our expectations are too optimistic, we still expect to get a huge amount of valuable data, including the discovery of a great many new exoplanets. The mission is a 'safe bet'. Suppose this is in competition with another mission, primarily designed to explore a relatively unknown parameter space. The more unknown the parameter space, the riskier the mission is perceived to be: it is likely to be thought of (disparagingly) as a high-risk 'fishing expedition'. An investor (funding body) has a simple choice between a near-zero-risk investment and a high-risk investment.[5]

---

[4] Cf. Cleland (2019a, p. 726): "It is very rare […] for an anomaly to be anticipated before it is encountered."

At any one time, certain minority groups in the community are researching non-mainstream – and thus intellectually risky – theoretical ideas. For example, some astrobiologists are researching non-mainstream biosignatures, e.g. signatures based on life that is not carbon-based. If we really do 'expect the unexpected', we should want to support such research, since we should expect that it really does have a non-negligible chance of paying off (and in a big way). But such research is perceived as 'high risk, high reward', and the term 'high risk' (much more than 'high reward') is going to be especially salient to potential funders. This is essentially because "bad is stronger than good" (Baumeister *et al*. 2001). Indeed, as recently as 2009 NASA rejected missions based solely on a risk assessment (Schingler *et al*., 2009) – high risk projects were rejected. An aversion to 'risky' projects is also in evidence in the public briefing[6] accompanying the publication of NASEM (2018); Dave Charbonneau (one of the two co-chairs of the committee) says:

> There's another branch of astrobiology that is looking for alternate biosignatures… very interesting… really *not* the driving force of the current strategy in terms of the telescopic observations. (Charbonneau's emphasis)

---

[5] We are of course talking here of what is sometimes called *intellectual* risk, as opposed to *technical* risk. For simplicity I assume that technical risk can always be made very low by elevating the cost. More on 'risk' below.
[6] https://livestream.com/NASEM/events/8339907/videos/179863839 (last accessed August 2020).

Obviously when he says that this minority sub-community "is looking for alternate biosignatures" he means with pen, paper, and computer simulations! They won't be looking for relevant biosignatures with high-spec instrumentation any time soon.

Thus there are two different ways in which 'the unexpected' might be missed for a very long time, or even indefinitely: (i) we have many targets we expect to be interesting, and these take up all the mission time (e.g. for JWST), meaning that the huge number of low-priority targets never get looked at, and (ii) we have many projects we expect to be fruitful, and these take up all the available funding, meaning that 'high risk' projects based on non-mainstream theory never get funded.

## 4. How to 'expect the unexpected' – adopt a balanced portfolio?

If we take these threats seriously, what can be done? One thought is that we simply have to base our resource allocation strategy on the best theoretical ideas we have right now, and we just have to accept that sometimes we'll miss something unexpected, potentially for a long time. But we needn't be so defeatist. We might instead be inspired by a popular financial investment strategy, where we adopt a 'balanced portfolio', striking a balance between higher risk (but also higher reward) investments, and lower risk (but lower reward) investments. Such an emphasis on balance is well-known in the literature on the division of cognitive labor in scientific communities. E.g. Weisberg and Muldoon (2009) provide an analysis indicating that, "a mixed strategy where some scientists are very conservative and others quite risk taking leads to the maximum amount of epistemic progress in the scientific community." (p. 227). Unfortunately, Weisberg and Muldoon's model applies poorly in a context where the risk-takers' research is inhibited by journal editors, funding bodies, and/or other community dynamics.[7] That is, we need to ensure that risk-takers are allowed to *be* risk-takers in the contemporary scientific context, where scientific communities are integrated to a greater extent than ever before, and scientists cannot be accurately modelled as isolated individuals (*contra* Strevens 2003).[8]

---

[7] Cf. Cleland (2019a, p. 724), discussing Carl Woese's work on the evolutionary significance of the prokaryote-eukaryote distinction: "Woese's work was not greeted with enthusiasm. Many biologists ignored it, and several eminent members of the biological community made fun of it."

Thus we want not merely a balance of low/medium/high-risk *research appetites* in the scientific community, but in addition a community dynamic where resources are allocated to reflect such a balance of appetites. Evidence that there is a mismatch between research appetites and resource allocation (see Fig. 1) comes in two forms:

(i)     research into the relevant community dynamics (Haufe 2013; Stanford 2019);

(ii)    concrete examples of unfortunately inhibited 'high risk' research from the history of science, including:

- continental drift theory 1915-1960;

- the bacterial theory of peptic ulcer disease 1954-1985 – see e.g. Zollman (2010).

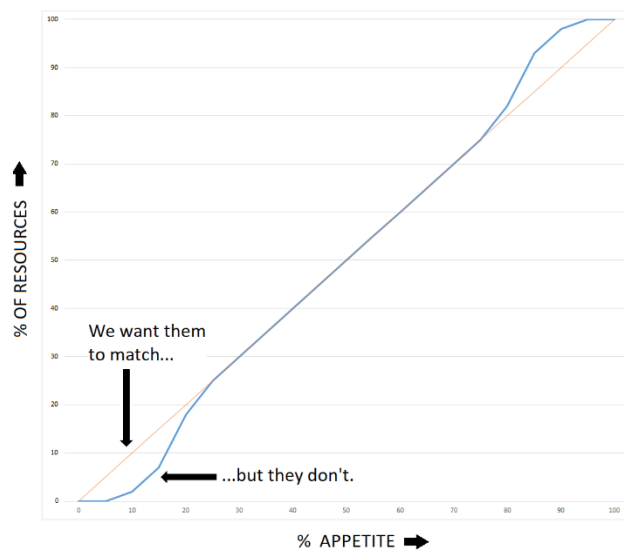

**Fig. 1.** The appetite-resources mismatch: at the extremes there is a mismatch between the percentage of the community with an appetite to

---

[8] Strevens (2003) writes, "with respect to worker hours at least, [resource] allocation in science is driven to a great extent by certain decisions of individual scientists, namely, their decisions as to what projects to pursue." (p. 64) But Stanford (2019) argues that contemporary scientists are not free to make such 'individual' decisions; natural risk-takers are under pressure to tow the line.

pursue a given line of research, and the percentage of resources made available for that line of research.

To give just one line of argument (cf. Stanford 2019, Section 1.3): if only 5% of the community think a hypothesis is worth pursuing, then if one such researcher applies for funding it is likely that the peer-reviewers and members of the funding committee will consider it a hypothesis that is *not* worth pursuing. If the researcher is lucky and one or two members of the funding committee consider the hypothesis worth pursuing, it is still very unlikely that the project will be funded; it will inevitably be competing with projects that *every* member of the funding committee considers eminently worthy of funding. Stanford isn't thinking about the space science context, but consider the following from a NASA white paper (Schingler 2009):

> Since proposing teams are aware of their audience and the TMC [risk assessment] process, teams take a more conservative approach and self-censor more innovative (and arguably riskier) approaches, even if they would save money or would provide a much higher scientific yield.[9]

If this is right, and non-mainstream research judged to be 'high risk' by 95% (say) of the relevant scientific community is significantly inhibited, then the community is raising a barrier between its practices and 'the unexpected', especially when we think about particularly *big* surprises that are most likely out there waiting to be discovered (Fig.2).

---

[9] Developments at NASA since 2009 will be discussed in Section 6, below.

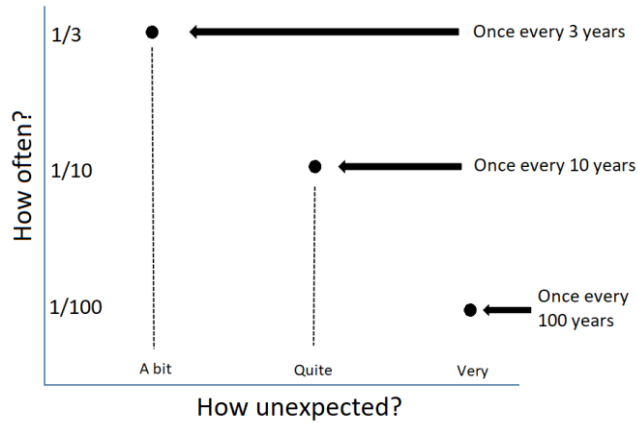

**Fig. 2.** A simple model of the frequency and magnitude of surprises in science.

This all adds to the case that we cannot be laid-back about resource allocation; the status quo is not optimum. And why would it be? As Kitcher (1990, p. 22) remarks: "[I]t would be highly surprising if the existing social structures of science … were to be vindicated by an optimality analysis." Scientists – as human beings more generally – are naturally risk averse. Kitcher (p. 7) cites Wegener's theory of continental drift and notes, "we may be grateful to the stubborn minority who continue to advocate problematic ideas."[10] The greater the integration of science, the more power the conservatives have to inhibit the research appetite of the small minority of risk-takers. Greater integration also increases the chance of a problematic 'bandwagon effect'; cf. Zollman (2010) considering the case of peptic ulcer disease: "Palmer's study was too influential. […] It was the widespread acceptance of Palmer's result which led to the premature abandonment of the diversity in scientific effort present a few years earlier." (p. 21). And the international scientific community is today more integrated than at any time in its history.

How would the 'balanced portfolio' proposal apply to a major new telescope such as JWST? We might adopt a policy to dedicate roughly 5% of its exoplanet time to 'high risk' targets, and roughly 20% of its time to targets judged to be 'medium risk', still leaving 75% of its time for mainstream, 'low risk' targets. Of course, there would be some difficult decisions to make, such as which projects to discount completely on the grounds that they are *too* risky for serious consideration. But these are hardly insuperable challenges.[11] Once the truly unpalatable proposals are eliminated, we might choose between the other available 'high risk' proposals at random. Or, if we really want to remove *all* theoretical expectations, we could forget about scientists' proposals, and devote one day in a hundred to a 'spin the bottle day' for JWST where a target planet which would otherwise never get looked at is chosen more or less at random.[12]

---

[10] Cleland's Woese example is also relevant here: "Undaunted, Woese continued his investigations and by 1990 was urging an even more radical restructuring of biological systematics […] It took almost 20 years for Woese's work to be widely accepted by biologists." (2019a, p. 724).

It is one thing to apply this 'balanced portfolio' thinking to observation targets; quite another to apply it to funding decisions generally. Space missions, and instruments, are often hugely expensive, so giving 5% of resources to a non-mainstream research program might not go very far. There is also the issue of how to judge between all the different non-mainstream options. Should 5% of resources be spread out between many different 'high risk' research programs? Or are we to choose the least risky high-risk project for the full 5% of funding? Or the *most* risky? There are bound to be *some* high-risk projects (e.g. life on Titan) where a dedicated satellite/mission or dedicated instrumentation would be needed – it wouldn't be a simple case of pointing our current telescope in a different direction. One option is for the relevant scientific sub-community to save up money for a great many years, until such a time as the required funds are available. Alternatively the sub-community in question could choose to devote these funds to theoreticians, with the thought that, once the theory is sufficiently robust, the corresponding proposals would move from 'high risk' to 'medium risk', thus opening up new sources of funding.

---

[11] The GSA (Geological Society of America) annual conference includes a session entitled 'Unconventional Ideas and Outrageous Hypotheses', precisely in an attempt to support high risk projects. Behind the scenes there have sometimes been disagreements concerning which submissions to this session were really *too* outrageous to be taken seriously.
[12] For JWST, only one day in five will be spent on exoplanet science, so 5% of its exoplanet time would amount to one day in a hundred.

## 5. Factoring in 'low risk, high reward' exploratory science

A 'spin the bottle day' for JWST may strike the reader as highly controversial. Many different scientists around the world have spent years researching precisely what they might do with JWST if they could just get their hands on it for a few days. We can imagine how these scientists would react to a 'spin the bottle day' for JWST (even just one day in a hundred) where random high risk targets are selected instead of their own carefully chosen ones. A 'balanced portfolio' strategy takes time away from well-thought-through targets, preferring targets that have not been thought through at all. Scientists have sometimes said that they like the idea of the community adopting a balanced portfolio; what could be wrong with 'balance'? But whilst it's easy to agree to this in the abstract, it's harder to swallow when one makes the suggestion concrete, as when we propose spending precious JWST time on high risk targets.[13] It might be urged that some such high risk targets will deliver unexpectedly high rewards. But what of the thought that we'll already get plenty of high rewards from the *low* risk targets?

Here we meet a disanalogy with financial investment 'balanced portfolio' strategies. Financial investors almost never have an opportunity for low risk, high reward investments; their sole motivation for pursuing high risk options lie with the fact that this is (usually) the only way to get high rewards. If they could get comparable high rewards with low risk options, they'd never take those high risks. But in the contemporary space science context, we *can* see before us low risk, high reward projects.

Take, for example, the LUVOIR telescope concept (cf. NASEM 2018, pp. 80 and 135). This is very low risk, since we know for sure (ignoring 'technical risk') that it would deliver all sorts of step-change scientific returns. Most significant of all, it would be able to directly image exoplanets, perhaps even imaging a 'pale blue dot' exoplanet similar to Earth that has a decent chance of being inhabited with at least primitive life. Taking everything into account, LUVOIR (though expensive) is a clear 'very low risk, very high reward' project. A direct image of an Earth-like exoplanet would immediately become one of the most influential images of all time, making the front page of every media outlet in the world.[14] There is no higher reward (within grasp) in the field of exoplanets, and we don't need anything like a 'high risk' project to go after this extraordinary return. Similarly with many other low risk, high reward mission concepts being placed in front of NASA and other space agencies right now.

---

[13] Cf. Maccrimmon and Wehrung (1988, p. 231): "Managers have a tendency to say they are risk takers, but in their actions they are more likely to be risk averters."

This might be called the 'low hanging fruit argument'. If the apples hanging low are just as juicy as those higher up, we have no reason to put *any* effort into reaching those apples that are difficult to reach. In the context of contemporary space science, this would mean leaving high risk projects behind, at least for now – not even devoting 5% of resources to them. This seems backward when we think about cases where inhibiting the non-mainstream/high risk projects turned out to be a big mistake. However, not all scientific contexts are alike. Those historical cases where the minority view was unfortunately inhibited are cases where scientists were adjudicating between competing theories. In the current space science context scientists are not adjudicating between opposing theories in anything like the same way. Instead, exoplanet science is sensibly characterised as *exploratory science* – the exploration of a huge range phenomena that has recently become accessible. And we know enough about the relevant parameter space to know that major scientific rewards are out there, in abundance, waiting to be claimed.

We can make this more concrete by considering observation target options for JWST. Three of the debated options are:

i.  Gas giant planets and mini-Neptunes. JWST would be good at analysing these planets, and could look at a relatively large number of them, such that a good cache of results would be guaranteed. In the process we'd expect to learn a lot about the formation of planetary

---

[14] The NASEM strategy document imagines "an ephocal moment in human history", when we acquire "evidence of a habitable Earth twin." (p. 78).

systems. Specific targets could be GJ 436b and GJ 1214b. (NASEM 2018, p. 86; Kempton *et al.* 2018)

ii. Rocky planets, such as those within the TRAPPIST-1 system. This would be much more time-intensive; far fewer planets could be analysed in the same amount of time. However, many scientists are excited to learn more about the TRAPPIST-1 system. (Lustig-Yaeger et al. 2019)

iii. K2-22b, a disintegrating rocky planet orbiting its parent star once every nine hours. JWST would be able to analyse the disintegration (via transmission spectroscopy of the planet's dust tail), thereby providing information on the interior of the planet. This may be a rare opportunity to 'see' the interior of a rocky planet. (Bodman *et al.* 2018)

These options are all low risk, in the sense that it is vanishingly unlikely that they would deliver no results of interest. And they all offer potentially quite high rewards: the information we would expect to acquire in each case would be extremely valuable. If we are pushed to decide between them, one reasonable answer seems to be that there just is no right answer here. Instead we are faced with a mere difference of tastes, preferences, or values. In which case the decision might be made by a simple 'show of hands' (or a proxy).

However, option (ii) does stand out, in that it offers the possibility of a *very* high reward in a way (i) and (iii) don't. If we turn to the planets in the TRAPPIST-1 system, there is a non-negligible chance that one or more of these planets is inhabited. The probability of this is judged by many to be quite low, given the aggressive activity of the TRAPPIST-1 host star (past and present), and the fact that these planets are probably tidally locked to the star. However, relevant theory is tentative, and the majority of scientists agree that we can put only limited trust in any given model[15]; we really just need to go and look at the planets and find out. Certainly we know that the TRAPPIST-1 system contains small rocky planets in the 'classical' habitable zone, suitable for surface liquid water.

---

[15] E.g. Dong et al. (2018); Dencs and Regaly (2019). As Victoria Meadows has recently said of the TRAPPIST-1 system, "We're ready to be amazed by what we find." https://www.ustream.tv/playlist/590419/video/124551395 (last accessed August 2020).

Thus option (ii) is more clearly 'low risk, high reward' than options (i) and (iii), chiefly because when we think about the 'high rewards' that are reasonable to hope for from options (i)-(iii), option (ii) stands out. Whilst exoplanet scientists do indeed have different values and preferences, one thing unites them: the search for extraterrestrial life. Scientists are united when it comes to the high value of discovering (even primitive) life.

The promise of this particular high reward from option (ii) could disappear very quickly. JWST may discover relatively quickly that none of the TRAPPIST-1 planets have substantial atmospheres (never mind *interesting* atmospheres), and are all no more habitable than modern-day Mars. Such a discovery would immediately reduce the value of pursuing an analysis of these planets. Thus a balanced approach to options (i)-(iii) might go as follows:

> Pursue option (ii); if atmospheres are identified continue the analysis, but if there are no atmospheres move on to options (i) and (iii).

Note how such a strategy goes firmly against the 'spin the bottle' idea mentioned earlier. With these options before us, it just isn't reasonable to spend any JWST time (even 5%) on targets we currently have no good reason to think are interesting.

## 6. Developments at NASA since 2009, and a proposal

A concern remains: according to a NASA white paper (Schingler *et al.* 2009), as recently as 2009 proposals were rejected if they were judged to be 'high risk', regardless of any other considerations. The foregoing argument suggests that this was a *good* strategy: high risk proposals aren't worth looking at, since we can get high rewards from low (and of course medium) risk proposals. But surely throwing all high risk proposals in the bin was too crude?

Developments at NASA since 2009 indicate that this has indeed been judged as too crude. First, a policy was put in place to simply flag high risk high impact proposals for further scrutiny. Then, in a 2017 review of NASA's Planetary Science Division, the National Academies of Sciences, Engineering, and Medicine explicitly put forth the following recommendation:

> *Recommendation*: NASA needs to investigate appropriate mechanisms to ensure that high-risk/high-payoff fundamental research and advanced technology-development activities receive appropriate consideration during the review process. (NASEM 2017, p. 31)

This led to a survey of peer reviewers, for 1,577 proposals submitted to the 2017 Research Opportunities in Space and Earth Science (ROSES) funding competition. Reviewers were provided with definitions of 'high impact' and 'high risk', and asked to flag those proposals coming out as both high risk and high impact. 10% of proposals were flagged, and out of these 35% were selected for funding, compared with an overall selection rate of 24%. This *seemed* to show that there is in fact no problem with the funding of high risk high impact projects within NASA; Paul Hertz – Director of the Astrophysics Division at the Science Mission Directorate – stated in his October 22, 2018, NASA Astrophysics Update that this 'disproves the myth' that there is a problem with the funding of such projects.[16]

Given all the reasons for supposing that there would be natural reticence to both submit and fund such projects, together with the point made above that many projects can be high impact without needing to be high risk, this seems like an odd result.[17] Has something gone wrong? The definition of 'risk' used for the 2017 survey (and still in place in January 2020) was:

---

[16] Paul Hertz's presentation can be accessed here: https://science.nasa.gov/researchers/nac/science-advisory-committees/apac (see slide 4, and slides 23-25; last accessed August 2020.) He also presented on this on February 12, 2019; see here: https://smd-prod.s3.amazonaws.com/science-red/s3fs-public/atoms/files/Hertz_AAS_2019-Feb_Virtual_TH_Final.pdf (see slide 31; last accessed August 2020).

[17] For further discussion on the aversion to funding high risk projects see National Research Council (2010, p. 41), which directly inspired the NASEM (2017) recommendation, and Institute of Medicine (2007, pp. 149-50), which recommends devoting 8% of federal research agency funds "to catalyze high-risk, high-payoff research."

> To what extent would this proposal test novel or significant hypotheses, for which there
>
> is scant precedent or preliminary data or which run counter to the existing scientific
>
> consensus?

Reviewers had to select: (i) great extent, (ii) some extent, or (iii) little or none, with 'great extent' corresponding to 'high risk'. But this definition is too easy to meet.[18] There are many projects which test 'significant hypotheses', and for which there is 'scant precedent', but where, for theoretical reasons, we fully expect the project to deliver exciting results. And a project we fully expect to deliver exciting results is not a high risk project. Consider for instance the WFIRST microlensing survey and the Rosetta mission to land a module (*Philae*) on a comet. A reviewer might think they both fit the definition, since they both 'test novel or significant hypotheses', and for both there is 'scant precedent'. However, although there is indeed 'scant precedent' for these projects, they are both low risk, since on theoretical grounds we fully expect(ed) significant results (supposing nothing goes wrong *technically*, and of course the *Philae* project didn't go to plan; it was arguably risky in the technical sense, but that should be irrelevant).

Thus the worry is that (far) less than 10% of the submitted proposals were truly high risk, because reviewers judged them to meet the given definition for 'high risk' whereas, in fact, there was little (intellectual) risk involved. This hypothesis could be tested, by looking at the 54 relevant projects (34% of 10% of 1,577 submitted proposals), and seeing how many of them ultimately fail(ed) to deliver significant scientific returns. If they are all truly high risk, then a good number of them *should* fail. Alternatively, one could test the definition by handing descriptions of twenty past projects (say) to a number of space scientists, and asking them the following questions: (i) Did this project test novel or significant hypotheses? (ii) Was there 'scant precedent' for this project? If they answer 'yes, to a great extent' for projects we are sure should *not* come out as 'high risk' in the intellectual sense, then that would reveal a problem with the definition.

---

[18] Another problem is that it assumes projects will be testing a hypothesis but, as already argued, the contemporary exoplanet field is full of 'exploratory science', where one is not (at least not primarily) testing a hypothesis.

The point here is not to argue that there really *is* a problem with funding high risk high impact projects within space science. Rather, the point is that we shouldn't fool ourselves that such projects are being regularly submitted and funded if they are not. If they are *not* being regularly submitted and funded, there might be good reasons for that. For example, as noted in the previous section, given all the exciting fully mainstream targets for JWST, we wouldn't expect scientists to request funding for a high risk, theoretically fringe target and, if such a funding proposal *were* made, it would be very reasonable to reject it. To put it even more starkly, if 0% of target proposals for JWST are high risk, or if 0% of high risk proposals get funded, that need *not* indicate a *problematic* reticence to fund high risk high impact projects. Rather, it might be the only reasonable thing to do in a context where there are several low risk high impact options.

Should we go back, then, to the situation within NASA prior to 2009, where all high risk proposals were immediately rejected? Surely not. It has been here argued that scientific communities sometimes *do* struggle to support high risk projects, even to the extent that projects crucial for scientific progress and the pursuit of truth (e.g. continental drift theory) were seriously inhibited for many years. What we really want is a system that allows high risk projects to be fairly judged alongside low and medium risk projects.

Schingler *et al.* (2009) briefly proposes "a mission selection system based on the following Mission Evaluation Metric:

$$V = S \times P_S / \$$$

…where S = science value, $P_S$= Probability of Success of the mission and \$ is the cost." This is a good initial model in the sense that it allows for various balances between risk, reward, and cost to come out with the same score. It also allows for high risk high reward proposals to score *higher* than low risk high reward proposals, if the cost of the former is small compared with the latter. Thus, just because there are low risk high reward options, that doesn't mean that all high risk proposals should be immediately dismissed.

This proposal is certainly just a starting point. Suppose that various projects put forward for funding are scored according to the given Mission Evaluation Metric; we might *not* want to simply award funding to the highest scorers, since that is consistent with a situation where one incredibly expensive project takes all the funding, and there could reasonably be objections to that. In addition, the highest scoring proposals could *all* be 'high risk', and this would be worrying, since one would then expect a good number of the projects to fail, and that would be embarrassing from a public and political point of view. It is important that relatively few missions are seen to 'fail', since there is a great tendency to perceive such projects as a waste of money:

> [H]igh-risk projects are prone to failing and increased government and public scrutiny make "projects that fail" increasingly untenable. (National Research Council 2010, p. 41)

If many high-scoring proposals *were* high risk, it would seem reasonable to adopt a balanced portfolio of low/medium/high risk projects, even if this would mean funding some low risk projects that scored lower than some high risk projects. However, this is merely hypothetical since, as already noted, in the contemporary space science context we might well find that no high risk proposals are competitive, even when they offer high rewards.

A note on the definition of 'high risk': In the Schingler et al. (2009) model 'high risk' is clearly associated with "low probability of success of the mission". This is an improvement on the 'scant precedent' definition used recently within NASA, but still by no means ideal. A project to explore a little-known parameter space where the scientific return can't be predicted in advance ought to come out as 'high risk' – the project might deliver little of much scientific interest.[19] And yet, the mission would still have been successful in the sense that it successfully explored the parameter space it set out to explore. What we really need is a definition sensitive to the probability that there will be few/no significant scientific returns.[20] My suggestion, then, is that a project should be labelled 'high risk' just when the probability of a significant scientific return is significantly below average. In any future survey comparable to the ROSES-2017 survey, I recommend the following question be asked:

> Assuming the proposed project goes to plan, do you judge that the probability
> of significant scientific returns is (i) average/above average, (ii) below average,
> (iii) significantly below average?

It would be very interesting to compare the results of a new survey, asking this question, with the ROSES-2017 survey.

## 7. Bias reduction strategies in the search for extraterrestrial life

This paper has articulated ways in which theoretical biases can and do hinder scientific progress. We need concrete, implementable strategies for minimizing such biases. The primary strategy offered here is for funding bodies to be especially sensitive to the value of funding a balanced portfolio of projects, including high risk projects so long as:

(i)       they are not too expensive,

(ii)      the rewards are potentially transformative, and

(iii)     there isn't a low or medium risk route to the same/similar rewards.

---

[19] An example would be Gordon Walker's twelve-year search for exoplanets, 1980-1992, ending in a null result (Walker *et al*. 1995).
[20] The word 'significant' needs to be included, since every project will generate *some* scientific returns.

The previous section showed the care that is necessary when we define the term 'high risk'; most importantly, a significant percentage of high risk projects are expected to fail, in the sense that the final scientific payoff is of limited significance compared with initial ambitions. We must learn to accept such 'failures'; the only way to avoid them is to never fund high risk projects.

At several points already in this paper I have referenced Cleland (2019a, 2019b) who is also motivated by the fact that we need bias reduction strategies in the search for extraterrestrial life. Her own strategy is based on two specific concerns vis-à-vis the way astrobiologists approach phenomena: (i) they are too wedded to *definitions* of life, and (ii) they are slow to recognise when a phenomenon is *anomalous*, in the sense that it strongly resists any (good) explanation "within a framework of widely accepted scientific beliefs" (2019b, p. 180). Thus she advocates that scientists employ a suite of *tentative* (as opposed to defining) criteria for life in order to "identify phenomena that resist classification as living or nonliving as worthy of further investigation for novel life" (2019a, p. 722). The strategy is designed to "[expedite] the recognition that a phenomenon is anomalous" (2019b, p. 178).

Cleland's strategy and my own are starkly different, but not in the sense that they conflict; they are fully complementary. Cleland is motivated by examples where scientists' theoretical biases bring them to misinterpret, or at least too hastily interpret, a given phenomenon; she provides a list of such examples in Cleland (2019a, Section 2) and Cleland (2019b, Section 8.2). What is central to all of Cleland's examples is the way scientists are too quick to fit a puzzling phenomenon into the standard theoretical framework, and are too slow to recognize that what is in front of them is a genuine anomaly, potentially requiring important theoretical adjustments. As she writes in Cleland (2019b, p. 178), "My focus in this chapter is … on expediting the recognition that a phenomenon is anomalous," and all of her examples fit with that explicit focus.

Cleland's focus is different from the focus of this paper, and this serves to illustrate a very important point: theoretical biases are damaging to scientific progress in *more than one way*. My own focus throughout has been on cases where an unexpected phenomenon is completely missed, as in the case of the gorilla costume example, the anomalous playing card example, and the hot Jupiter example. Thus theoretical biases hinder progress in at least these two ways:

(i)     by blocking or delaying the discovery of unexpected phenomena;

(ii)    by blocking or delaying the recognition that a given phenomenon is anomalous.

There is no reason to suppose that there is just one way to respond to these different barriers to progress. Cleland's strategy is more focused on the way scientists approach the concept of 'life', how they think about biosignatures, and the difference between a merely 'puzzling' phenomenon and a genuine anomaly. My strategy is more focused on funding-body decisions, and the distribution of intellectual risk across the community. It is also broader, in the sense that it isn't restricted to the search for life, but concerns the funding of scientific projects more generally. Neither strategy demands a major overhaul of the status quo. But if both strategies are implemented this will be a significant step in the right direction, removing barriers to progress caused by theoretical biases, and conducive to the most surprising, and thus most significant, scientific discoveries.

## References

Baumeister RF, Bratslavsky E, Finkenauer C and Vohs KD (2001) Bad is stronger than good. *Review of General Psychology* **5**(4), 323-370.

Bodman EHL, Wright JT, Desch SJ and Lisse CM (2018) Inferring the composition of disintegrating planet interiors from dust tails with future James Webb Space Telescope observations. *The Astronomical Journal* **156**(4), 173 (8pp).

Boss A (2009) *The Crowded Universe*. New York: Basic Books.

Bruner JS and Postman L (1949) On the Perception of Incongruity: A Paradigm. *Journal of Personality* **18**, 206-223.

Cleland CE (2019a) Moving Beyond Definitions in the Search for Extraterrestrial Life. *Astrobiology* **19**(6), 722-729.

Cleland CE (2019b) *The Quest for a Universal Theory of Life: Searching for Life as We Don't Know It*. Cambridge: Cambridge University Press.

Dencs Z and Regály Zs (2019) Water Delivery to the TRAPPIST-1 Planets. *Monthly Notices of the Royal Astronomical Society* **487**(2), 2191–2199.

Dong C, Jin M, Lingam M, Airapetian VS, Ma Y and van der Holst B (2018) Atmospheric escape from the TRAPPIST-1 planets and implications for habitability. *Proceedings of the National Academy of Sciences of the United States of America* **115**(2), 260.

Gillon M, Triaud AHMJ, Demory BO, Jehin E, Agol E, Deck KM, Lederer SM, et al. (2017) Seven temperate terrestrial planets around the nearby ultracool dwarf star TRAPPIST-1. *Nature* **542**(7642), 456-460.

Goldsmith D (2018) *Hidden Worlds and the Quest for Extraterrestrial Life*. Cambridge, Mass.: Harvard University Press.

Hacking I (1983) *Representing and Intervening*. Cambridge, UK: Cambridge University Press.

Haufe C (2013) Why do funding agencies favor hypothesis testing? *Studies in History and Philosophy of Science* **44**(3), 363-374.

Huang CX, Shporer A, Dragomir D, Fausnaugh M, Levine AM, Morgan EH, Nguyen T, Ricker GR, Wall M, Woods DF and Vanderspek RK (2018) Expected yields of planet discoveries from the TESS primary and extended missions. *Earth and Planetary Astrophysics* **1807**, 11129.

Institute of Medicine (2007) *Rising Above the Gathering Storm: Energizing and Employing America for a Brighter Economic Future*. Washington DC: The National Academies Press. https://doi.org/10.17226/11463.

Joshi MM and Haberle RM (2012) Suppression of the water ice and snow albedo feedback on planets orbiting red dwarf stars and the subsequent widening of the habitable zone. *Astrobiology* **12**(1), 3.

Kempton EM-R, Bean JL, Louie DR, Deming D, Koll DDB, Mansfield M, Lopez-Morales M, *et al*. (2018) A framework for prioritizing the *TESS* planetary candidates most amenable to atmospheric characterization. *Publications of the Astronomical Society of the Pacific* **130**(993), 114401 (14pp).

Kitcher P (1990) The Division of Cognitive Labor. *The Journal of Philosophy* **87**(1), 5-22.

Kuhn T (1970) *The Structure of Scientific Revolutions*, 2nd edition. Chicago: University of Chicago Press.

Luger R and Barnes R (2015) Extreme water loss and abiotic O-2 buildup on planets throughout the habitable zones of M dwarfs. *Astrobiology* **15**(2), 119-143.

Lustig-Yaeger J, Meadows VS, Lincowski AP (2019) The Detectability and Characterization of the TRAPPIST-1 Exoplanet Atmospheres with JWST. *The Astronomical Journal* **158**(1), 27 (28pp).

Maccrimmon KR and Wehrung DA (1988) *Taking Risks: The management of uncertainty*. New York: Macmillan.

Marcy GW and Butler RP (1998) Detection of Extrasolar Giant Planets. *Annual Review of Astronomy and Astrophysics* **36**, 57–97.

Mayor M and Queloz D (1995) A Jupiter-mass companion to a solar-type star. *Nature* **378**, 355-359.

Moores JE, Atreya SK, Gough RV, Martinez GM, Meslin PY, *et al.* (2019) Methane Seasonal Cycles at Gale Crater on Mars Consistent with Regolith Adsorption and Diffusion. *Nature Geoscience* **12**, 321-325.

NASEM (2017) *Review of the Restructured Research and Analysis Programs of NASA's Planetary Science Division* (National Academies of Science, Engineering, and Medicine). Washington DC: The National Academies Press; https://www.nap.edu/catalog/24759/review-of-the-restructured-research-and-analysis-programs-of-nasas-planetary-science-division.

NASEM (2018) *Exoplanet Science Strategy* (September 2018, National Academies of Science, Engineering, and Medicine). Washington DC: The National Academies Press; https://sites.nationalacademies.org/SSB/CompletedProjects/SSB_180659.

National Research Council (2010) *An Enabling Foundation for NASA's Earth and Space Science Missions*. Washington DC: The National Academies Press; https://doi.org/10.17226/12822.

Oreskes N and Conway EM (2010) *Merchants of Doubt*. Bloomsbury Publishing USA.

Schingler R, Marshall W, MacDonald A, Lupisella M and Lewis B (2009) ROSI – Return on Science Investment: A system for mission evaluation based on maximizing science, a white paper in support of the *Planetary Science Decadal Survey 2013-2022*, a report of the National Research Council. Washington DC: National Academies Press.

Schwieterman EW, Meadows VS, Domagal-Goldman SD, Deming D, Arney GN, Luger R, Harman CE, Misra A and Barnes R (2016) Identifying planetary biosignature impostors: Spectral features of CO and O4 resulting from abiotic $O_2/O_3$ production. *Astrophysical Journal Letters* **819**(1), L13.

Simons DJ and Chabris CF (1999) Gorillas in our midst: sustained inattentional blindness for dynamic events. *Perception* **28**, 1059-1074.

Stanford K (2019) Unconceived alternatives and conservatism in science: the impact of professionalization, peer-review, and Big Science. *Synthese* **196**(10), 3915-3932.

Strevens M (2003) The Role of the Priority Rule in Science. *The Journal of Philosophy* **100**(2), 55-79.

Walker GAH (2003) Seeking other solar systems. Chapter Twenty in Garwin L and Lincoln T (eds), *A Century of Nature*: *Twenty-One Discoveries that Changed Science and the World*. Chicago: University of Chicago Press, pp. 313-332.

Walker GAH, Walker AR, Irwin AW, Larson AM, Yang SLS and Richardson DC (1995) A Search for Jupiter-Mass Companions to Nearby Stars. *Icarus* **116**, 359-375.

Weisberg M and Muldoon R (2009) Epistemic Landscapes and the Division of Cognitive Labor. *Philosophy of Science* **76**(2), 225-252.

Zollman KJS (2010) The Epistemic Benefit of Transient Diversity. *Erkenntnis* **72**, 17-35.